\documentclass{amsart}
\usepackage{amsfonts}
\usepackage{graphicx}
\usepackage{amscd}
\usepackage{amsmath}

\theoremstyle{plain}

\numberwithin{equation}{section}

\input{tcilatex}
\input tcilatex

\begin{document}
\title[Quantum Measurement Problem]{On the Dynamical Solution of Quantum
Measurement Problem}
\author{V P Belavkin}
\address{Mathematics Department, University of Nottingham, NG7 2RD, UK}
\email{vpb@maths.nott.ac.uk}
\thanks{Submitted to the Proceedings of International Conference\\
\textit{Quantum Theory: Reconsideration of Foundations --2}, Vaxjo 2003}
\date{1--7 June 2003, Vaxjo (Sweeden)}
\keywords{}

\begin{abstract}
The development of quantum measurement theory, initiated by von Neumann,
only indicated a possibility for resolution of the interpretational crisis
of quantum mechanics. We do this by divorcing the algebra of the dynamical
generators and the algebra of the actual observables, or \textit{be}ables.
It is shown that within this approach quantum causality can be rehabilitated
in the form of a superselection rule for compatibility of the past beables
with the potential future. This rule, together with the self-compatibility
of the measurements insuring the consistency of the histories, is called the
nondemolition, or causality principle in modern quantum theory. The
application of this rule in the form of the dynamical commutation relations
leads in particular to the derivation of the von Neumann projection
postulate. This gives a quantum stochastic solution, in the form of the
dynamical filtering equations, of the notorious measurement problem which
was tackled unsuccessfully by many famous physicists starting with Schr\"{o}%
dinger and Bohr.
\end{abstract}

\maketitle
\urladdr{http://www.maths.nott.ac.uk/personal/vpb/}

\section{Introduction\protect\medskip \protect\medskip}

\begin{quotation}
\textit{How wonderful we have met with a paradox, now we have some hope of
making progress} - Niels Bohr.
\end{quotation}

In this paper we present the main ideas of modern quantum measurement theory
and the author's views on the quantum measurement problem which might not
coincide with the present scientific consensus that this problem is
unsolvable in the standard framework, or at least unsolved \cite{BuLa91}. It
will be shown that there exists such solution along the line suggested by
the great founders of quantum theory Schr\"{o}dinger, Heisenberg and Bohr.
We shall see that von Neumann only partially solved this problem which he
studied in his Mathematical Foundation of Quantum Theory \cite{Neum32}, and
that the direction in which the solution might be found was envisaged by Scr%
\"{o}dinger \cite{Schr31} and J Bell \cite{Bell87}.

Recent phenomenological theories of continuous reduction, quantum state
diffusion and quantum trajectories extended the instantaneous projection
postulate to a certain class of continuous-in-time measurements. As we shall
see here, there is no need to supplement the usual quantum mechanics with
any of such generalized reduction postulate even in the continuous time.
They all have been derived from the time continuous unitary evolution for a
generalized Dirac type Schr\"{o}dinger equation with a singular scattering
interaction at the boundary of our Hamiltonian model, see the recent review
paper \cite{Be02}. The quantum causality as a new superselection rule
provides a time continuous nondemolition measurement in the extended system
which enables to obtain the quantum state diffusion and quantum trajectories
simply by time continuous conditioning called quantum filtering. Our
nondemolition causality principle, which was explicitly formulated in \cite%
{Be94}, admits to select a continuous diffusive classical process in the
quantum extended world which satisfies the nondemolition condition with
respect to all future of the measured system. And this allows\ us to obtain
the continuous trajectories for quantum state diffusion by simple filtering
of quantum noise exactly as it was done in the classical statistical
nonlinear filtering and prediction theory. In this way we derived \cite%
{Be79, Be80} the quantum state diffusion of a Gaussian wave packet as the
result of the solution of quantum prediction problem by filtering the
quantum white noise in a quantum stochastic Langevin model for the
continuous observation. Thus the \textquotedblleft
primary\textquotedblright\ for the conventional quantum mechanics stochastic
nonlinear irreversible quantum state diffusion appears to be the secondary,
as it should be, to the deterministic linear unitary reversible evolution of
the extended quantum mechanics containing necessarily infinite number of
auxiliary particles. However quantum causality, which defines the arrow of
time by selecting what part of the reversible world is related to the
classical past and what is related to the quantum future, makes the extended
mechanics irreversible in terms of the injective semigroup of the invertible
Heisenberg transformations induced by the unitary group evolution for the
positive arrow of time. The microscopic information dynamics of this event
enhanced quantum mechanics, or Eventum Mechanics, allows the emergence of
the decoherence and the increase of entropy in a purely dynamical way
without any sort of reservoir averaging.

Summarizing, we can formulate the general principles of the Eventum
Mechanics which unifies the classical and quantum mechanics in such a way
that there is no contradiction between the unitary evolution of the matter
waves and the phenomenological information dynamics such as quantum state
diffusion or spontaneous jumps for the events and the trajectories of the
particles. This is a conventional, \emph{non-stochastic} but time asymmetric
quantum mechanics in an extended Hilbert space, in which the true and hidden
observables, or beables are mathematically distiguished from the evolution
generators. It can be described by the following fitures:

\begin{itemize}
\item It is a reversible wave mechanics of the continuous unitary group
evolutions in an infinite-dimensional Hilbert space

\item It has conventional interpretation for the normalized Hilbert space
vectors\ as state-vectors ( probability amplitudes)

\item However not all operators, e.g. the dynamical generator (Hamiltonian),
are admissible as the potential observables

\item Quantum causality is statistical predictability of the quantum states
based on the results of the actual measurements

\item It implies the choice of time arrow and an initial state which
together with past measurement data defines the reality

\item The actual observables (beables) must be compatible with any operator
representing a potential (future) observable

\item The Heisenberg dynamics and others symmetries induced by unitary
operators should be algebraically endomorphic

\item However these endomorphisms form only a semigroup on the algebra of
all observables as they may be irreversible.
\end{itemize}

Note that the classical Hamiltonian mechanics can be also described in this
way by considering only the commutative algebras of the potential
observables. Each such observable is compatible with any other and can be
considered as an actual observable, or beable. However, the Hamiltonian
operator, generating a non-trivial Liouville unitary dynamics in the
corresponding Hilbert space, is not an observable, as it doesn't commute
with any observable which is not the integral of motion. Nevertheless the
corresponding Heisenberg dynamics, described by the induced automorphisms of
the commutative algebra, is reversible, and pure states, describing the
reality, remain pure, non disturbed by the measurements of its observables.
This is also true in the purely quantum mechanical case, in which the
Hamiltonian is an observable, as there are no events and nontrivial beables
in the conventional quantum mechanics. The only actual observables, which
are compatible with any \ Hermitian operator as a potential observable, are
the constants, i.e. proportional to the identity operator, as the only
operators, commuting with any such observable. Their measurements do not
bring new information\ and do not disturb the quantum states. However any
non-trivial classical--quantum Hamiltonian interactions cannot induce a
group of the reversible Heisenberg automorphisms but only a semigroup of
irreversible endomorphisms of the decomposable algebra of all potential
observables of the composed classical-quantum system. This follows from the
simple fact that any automorphism leaves the center of an operator algebra
invariant, and thus induces the autonomous noninteracting dynamics on the
classical part of the semi-classical system. This is the only reason which
is responsible for failure of all earlier desperate attempts to build the
reversible, time symmetric Hamiltonian theory of classical-quantum
interaction which would give a dynamical solution of the quantum decoherence
and measurement problem along the line suggested by von Neumann and Bohr.
There is no nontrivial reversible classical-quantum mechanical interaction,
but as we have seen, there is a Hamiltonian irreversible interaction within
the time asymmetric Eventum Mechanics.

The unitary solution of the described boundary value problem indeed induces
endomorphic semi-classical Hamiltonian dynamics, and in fact is underlying
in any phenomenological reduction model \cite{Be02}. Note that although the
irreversible Heisenberg endomorphisms of eventum mechanics, induced by the
unitary propagators, are injective, and thus are invertible by completely
positive maps, and are not mixed, they mix the pure states over the center
of the algebra. Such mixed states, which are uniquely represented as the
orthogonal mixture over the `hidden' variables (beables), can be filtered by
the measurement of the actual observables, and this transition from the
prior state corresponding to the less definite (mixed) reality to the
posterior state, corresponding to a more definite (pure) reality by the
simple inference is not change the reality. This is an explanation, in the
pure dynamical terms of the eventum mechanics, of the emergence of the
decoherence and the reductions due to the measurement, which has no
explanation in the conventional classical and quantum mechanics.

Our mathematical formulation of the eventum mechanics as the extended
quantum mechanics equipped with the quantum causality to allow events and
trajectories in the theory, is just as continuous as Schr\"{o}dinger could
have wished. However, it doesn't exclude the jumps which only appear in the
singular interaction picture, which are there as a part of the theory, not
only of its interpretation. Although Schr\"{o}dinger himself didn't believe
in quantum jumps, he tried several times, although unsuccessfully, to obtain
the continuous reduction from a generalized, relativistic, ``true Schr\"{o}%
dinger'' equation. He envisaged that `if one introduces two symmetric
systems of waves, which are traveling in opposite directions; one of them
presumably has something to do with the known (or supposed to be known)
state of the system at a later point in time' \cite{Schr31}, then it would
be possible to derive the `verdammte Quantenspringerei' for the opposite
wave as a solution of the future-past boundary value problem. This desire
coincides with the ``transactional'' \ attempt of interpretation of quantum
mechanics suggested in \cite{Crm86} on the basis that the relativistic wave
equation yields in the nonrelativistic limit two Schr\"{o}dinger type
equations, one of which is the time reversed version of the usual equation:
`The state vector $\psi $ of the quantum mechanical formalism is a real
physical wave with spatial extension and it is identical with the initial
``offer wave'' of the transaction. The particle (photon, electron, etc.) and
the collapsed state vector are identical with the completed transaction.' \
There was no proof of this conjecture, and now we know that it is not even
possible to derive the quantum state diffusions, spontaneous jumps and
single reductions from models involving only a finite particle state vectors 
$\psi \left( t\right) $ satisfying the conventional Schr\"{o}dinger equation.

Our new approach, based on the exactly solvable boundary value problems for
infinite particle states described in this paper, resolves the problem
formulated by Schr\"{o}dinger. And thus it resolves the old problem of
interpretation of the quantum theory, together with its infamous paradoxes,
in a constructive way by giving exact nontrivial models for allowing the
mathematical analysis of quantum observation processes determining the
phenomenological coupling constants and the reality underlying these
paradoxes. Conceptually it is based upon a new idea of quantum causality
called the nondemolition principle \cite{Be94} which divides the world into
the classical past, forming the consistent histories, and the quantum
future, the state of which is predictable for each such history.

Here we develope the discrete time approach introduced in \cite{Be79, Be01b}
for solving the famous Schr\"{o}dinger's cat paradox. We shall see that even
the most general quantum decoherence and wave packet reduction problem for
an instantaneous or even sequential measurements can be solved in a
canonical way which corresponds to adding a single initial cat's state. The
discrete time dynamical model used for this solution is in parallel with the
quantum stochastic model for continuous in time measurements suggested in 
\cite{Be89a, Be89b}, see also \cite{GoGr94, GGH95}. These models give the
dynamical justification of the projection and other phenomenological
decoherence and reduction postulates. They resolve the Schr\"{o}dinger cat
paradoxes of quantum measurement theory in a constructive way, giving exact
nontrivial models in the differential form of evolution equations for the
statistical analysis of quantum observation processes determining the
reality underlying these paradoxes. Conceptually they are based upon a new
idea of quantum causality as a superselection rule called the Nondemolition
Principle \cite{Be94} which divides the world into the classical past,
forming the consistent histories, and the quantum future, the state of which
is predictable for each such history. This new postulate of the modern
quantum theory making \ the solution of quantum measurement possible can not
be contradicted by any experiment as we prove that any sequence of usual,
``demolition'' measurements based on the projection postulate or any other
phenomenological measurement theory is statistically equivalent, and in fact
can be dynamically realized as a simultaneous nondemolition measurement in a
canonically extended infinite semi-quantum system. The nondemolition models
give exactly the same predictions as the orthodox, ``demolition'' theories,
but they do not require the projection or any other postulate replaced by
the nondemolition causality principle. We examine also the implications for
time reversibility and time arrow which follow from the quantum causality
principle.

\section{Generalized reduction and its dilation}

Von Neumann's measurement theory postulates the process of decoherence for
any wave-function $\psi \left( x\right) $ as an instantaneous transition, or
jump $\psi \psi ^{\dagger }\mapsto \rho $ into the mixture 
\begin{equation*}
\rho =\sum_{y}F\left( y\right) \psi \psi ^{\dagger }F\left( y\right)
=\sum_{y}\psi _{y}\psi _{y}^{\dagger }\Pr \left\{ y\right\}
\end{equation*}
of the eigen-functions $F\left( y\right) \psi $ of a discrete-spectrum
observable $\mathrm{Y}=\sum yF\left( y\right) $. Here $F\left( y\right) $ is
a complete orthogonal family of eigen-projectors 
\begin{equation*}
\mathrm{Y}F\left( y\right) =yF\left( y\right) ,\quad F\left( y\right)
^{2}=F\left( y\right) =F\left( y\right) ^{\dagger },\quad \sum F\left(
y\right) =\mathrm{I}
\end{equation*}
for the observable $\mathrm{Y}$ defining the\emph{\ posterior state vectors} 
$\psi _{y}=F\left( y\right) \psi /\left\| F\left( y\right) \psi \right\| $
for all measurement results $y$ which have nonzero probabilities $\Pr
\left\{ y\right\} =\left\| F\left( y\right) \psi \right\| ^{2}$. Note that
the projections $\psi _{1}\left( y\right) =F\left( y\right) \psi $ are
normalized as 
\begin{equation*}
\sum_{y}\left\| \psi _{1}\left( y\right) \right\| ^{2}=\sum_{y}\int \left|
\psi _{1}\left( x,y\right) \right| ^{2}\mathrm{d}\lambda _{x}=1
\end{equation*}
if $\left\| \psi \right\| ^{2}:=\int \left| \psi \left( x\right) \right| ^{2}%
\mathrm{d}\lambda _{x}=1$ with respect to a given (discrete or continuous)
measure $\lambda $ on $x$. According to the L\"{u}dger's projection
postulate \cite{Lud51}, the renormalized non-linear versions $\psi \mapsto
\psi _{y}$ of the linear transformations $\psi \mapsto \psi _{1}\left(
y\right) $ defines the new states after the measurement corresponding to the
measurement results $y$ (with $\Pr \left\{ y\right\} \neq 0$).

Obviously the projection postulate is only a phenomenological principle
which is inconsistent with the Schr\"{o}dinger's unitary evolution, and
therefore it requires a dynamical justification. There have been innumerous
attempts to derive the decoherence and the projection postulate as a sort of
approximation corresponding to a limiting procedure in a dynamical model of
the measurement-apparatus interaction. While it is in principle possible to
obtain the decoherence as the result of averaging with respect to the
additional (reservoir) degrees of freedom, any attempt to derive the
projection postulate faces the problem of applying it on a higher level.
Surely the nonexistence of the solution for a physically well-defined
problem simply indicates an incorrectness of its mathematical setting. It
was pointed out by Niels Bohr that it is not possible to resolve this
problem unless as the reservoir is considered dynamically as quantum but
statistically as classical system. Following this old idea we shall
formulate the measurement problem as a mathematical problem which has at
least one exact solution. This solution might be not most satisfactory for
physics, however it gives the idealized dynamical model for any quantum
sequential measurement, not only with discrete but also with continuous
spectra. Let us therefore describe the generalized instantaneous reduction
principle which includes the indirect measurements with continuous data.

The generalized reduction of the wave function $\psi \left( x\right) $,
corresponding to a complete measurement with discrete or continuous spectrum
of $y$, is described by a function $V\left( y\right) $ whose values are
linear operators $\mathfrak{h}\ni \psi \mapsto V\left( y\right) \psi $ for
each $y$ which are not assumed to be unitary on the quantum system Hilbert
space $\mathfrak{h}$, $V\left( y\right) ^{\dagger }V\left( y\right) \neq I$,
but have the following normalization condition. The resulting wave-function 
\begin{equation*}
\psi _{1}\left( x,y\right) =\left[ V\left( y\right) \psi \right] \left(
x\right)
\end{equation*}
is normalized with respect to a given measure $\mu $ on $y$ in the sense 
\begin{equation*}
\iint \left| \left[ V\left( y\right) \psi \right] \left( x\right) \right|
^{2}\mathrm{d}\lambda _{x}\mathrm{d}\mu _{y}=\int \left| \psi \left(
x\right) \right| ^{2}\mathrm{d}\lambda _{x}
\end{equation*}
for any probability amplitude $\psi $ normalized with respect to a measure $%
\lambda $ on $x$. This can be written as the isometry condition $\mathrm{V}%
^{\dagger }\mathrm{V}=\mathrm{I}$ of the operator $\mathrm{V}:\psi \mapsto
V\left( \cdot \right) \psi $ in terms of the integral 
\begin{equation}
\int_{y}V\left( y\right) ^{\dagger }V\left( y\right) \mathrm{d}\mu _{y}=%
\mathrm{I},\quad \mathrm{or}\quad \sum_{y}V\left( y\right) ^{\dagger
}V\left( y\right) =\mathrm{I}.  \label{4.1}
\end{equation}
with respect to the base measure $\mu $ which is usually the counting
measure, $\mathrm{d}\mu _{y}=1$ in the discrete case, e.g. in the case of
the projection-valued $V\left( y\right) =F\left( y\right) $. The general
case of orthoprojectors $V\left( y\right) =F\left( y\right) $ corresponds to
the Kr\"{o}nicker $\delta $-function $V\left( y\right) =\delta _{y}^{\mathrm{%
X}}$ of a self-adjoint operator $\mathrm{X}$ on $\mathfrak{h}$ with the
discrete spectrum coinciding with the measured values $y$.

As in the simple example of the Schr\"{o}dinger's cat the dynamical
realization of such $V$ can always be constructed in terms of a unitary
transformation on an extended Hilbert space $\mathfrak{h}\otimes \mathfrak{g}
$ and a normalized wave function $\chi ^{\circ }\in \mathfrak{g}$. It is
easy to find such unitary dilation of any reduction family $V$ of the form 
\begin{equation}
V\left( y\right) =\mathrm{e}^{-i\mathrm{E}/\hbar }\mathrm{\exp }\left[ -%
\mathrm{X}\frac{\mathrm{d}}{\mathrm{d}y}\right] \varphi \left( y\right) =%
\mathrm{e}^{-i\mathrm{E}/\hbar }F\left( y\right) ,  \label{4.2}
\end{equation}%
given by a normalized wave-function $\varphi \in L^{2}\left( \mathbb{G}%
\right) $ on a cyclic group $\mathbb{G}\ni y$ (e.g. $\mathbb{G}=\mathbb{R}$
or $\mathbb{G}=\mathbb{Z}$). Here the shift $F\left( y\right) =\varphi
\left( y-\mathrm{X}\right) $ of $\chi ^{\circ }=\varphi $ by a measured
operator $\mathrm{X}$ in $\mathfrak{h}$ is well-defined by the unitary
shifts $\mathrm{\exp }\left[ -x\frac{\mathrm{d}}{\mathrm{d}y}\right] $ in $%
\mathfrak{g}=L^{2}\left( \mathbb{G}\right) $ in the eigen-representation of\
any selfadjoint $\mathrm{X}$ having the spectral values $x\in \mathbb{G}$,
and $\mathrm{E}=\mathrm{E}^{\dagger }$ is any free evolution action after
the measurement. As was noted by von Neumann for the case $\mathbb{G}=%
\mathbb{R}$\ in \cite{Neum32}, the operator $\mathrm{S}=\mathrm{\exp }\left[
-\mathrm{X}\frac{\mathrm{d}}{\mathrm{d}y}\right] $ is unitary in $\mathfrak{h%
}\otimes \mathfrak{g}$, and it coincides on $\psi \otimes \varphi $ with the
isometry $\mathrm{F}=\mathrm{S}\left( \mathrm{I}\otimes \varphi \right) $ on
each $\psi \in \mathfrak{h}$ such that the unitary operator $\mathrm{W}=%
\mathrm{e}^{-i\mathrm{E}/\hbar }\mathrm{S}$ \ dilates the isometry $\mathrm{V%
}=\mathrm{e}^{-i\mathrm{E}/\hbar }\mathrm{F}$ in the sense 
\begin{equation*}
\mathrm{W}\left( \psi \otimes \chi ^{\circ }\right) =\mathrm{e}^{-i\mathrm{E}%
/\hbar }\mathrm{S}\left( \psi \otimes \varphi \right) =\mathrm{e}^{-i\mathrm{%
E}/\hbar }\mathrm{F}\psi ,\quad \forall \psi \in \mathfrak{h}.
\end{equation*}%
The wave function $\chi ^{\circ }=\varphi $ defines the initial probability
distribution $\left\vert \varphi \left( y\right) \right\vert ^{2}$ of the
pointer coordinate $y$ which can be dispersionless only if $\varphi $ is an
eigen-function of the pointer operator $Y=\hat{y}$ (multiplication operator
by $y$ in $\mathfrak{g}$) corresponding to a discrete spectral value $%
y^{\circ }$ as a predetermined initial value of the pointer, $y^{\circ }=0$
say. This corresponds to ortho-projectors $V\left( y\right) =\delta _{y}^{%
\mathrm{X}}=F\left( y\right) $ ($\mathrm{E}=\mathrm{O}$) indexed by $y$ from
a discrete cycle group, $y\in \mathbb{Z}$ for the discrete $\mathrm{X}$
having eigenvalues $x\in \mathbb{Z}$ say. Thus the projection postulate is
always dilated by such shift operator $\mathrm{S}$ with $\chi ^{\circ
}\left( y\right) =\delta _{y}^{0}$ given as the eigen-function $\varphi
\left( y\right) =\delta _{y}^{0}$ corresponding to the initial value $y=0$
for the pointer operator $Y=\hat{y}$ in $\mathfrak{g}=L^{2}\left( \mathbb{Z}%
\right) $ \cite{StBe96}\ (In the case of the Schr\"{o}dinger's cat $\mathrm{U%
}$ was simply the shift $\mathrm{W}$ (mod 2) \ in $\mathfrak{g}=L^{2}\left(
0,1\right) :=\mathbb{C}^{2}$).

There exist another, canonical construction of the unitary operator $\mathrm{%
W}$ with the eigen-vector $\chi ^{\circ }\in \mathfrak{g}$ for a `pointer
observable' $Y$ in an extended Hilbert space $\mathfrak{g}$ even if $y$ is a
continuous variable of the general family $V\left( y\right) $. More
precisely, it can always be represented on the tensor product of the system
space $\mathfrak{h}$ and the space $\mathfrak{g}=\mathbb{C}\oplus L_{\mu
}^{2}$ of square-integrable functions $\chi \left( y\right) $ defining also
the values $\chi \left( y^{\circ }\right) \in \mathbb{C}$ at an additional
point $y^{\circ }\neq y$ corresponding to the absence of a result $y$ and $%
\chi ^{\circ }=1\oplus 0$ such that 
\begin{equation}
\langle x|V\left( y\right) \psi =\left( \langle x|\otimes \langle y|\right) 
\mathrm{W}\left( \psi \otimes \chi ^{\circ }\right) ,\quad \forall \psi \in 
\mathfrak{h}  \label{4.3}
\end{equation}
for each measured value $y\neq y^{\circ }$.

Now we prove this unitary dilation theorem for the general $V\left( y\right) 
$ by the explicit construction of the matrix elements $\mathrm{W}_{y^{\prime
}}^{y}$ in the unitary block-operator $\mathrm{W}=\left[ \mathrm{W}%
_{y^{\prime }}^{y}\right] $ defined as $\left( \mathrm{I}\otimes \langle
y|\right) \mathrm{W}\left( \mathrm{I}\otimes |y^{\prime }\rangle \right) $
by 
\begin{equation*}
\psi ^{\dagger }\mathrm{W}_{y^{\prime }}^{y}\psi ^{\prime }=\left( \psi
^{\dagger }\otimes \langle y|\right) \mathrm{W}\left( \psi ^{\prime }\otimes
|y^{\prime }\rangle \right) ,
\end{equation*}
identifying $y^{\circ }$ with $0$ (assuming that $y\neq 0$, e.g. $y=1,\ldots
,n$). We shall use the short notation $\mathfrak{f}=L_{\mu }^{2}$ for the
functional Hilbert space on the measured values $y$ and $\chi ^{\circ
}=|y^{\circ }\rangle $ (=$|0\rangle $ if $y^{\circ }=0$) for the additional
state-vector $\chi ^{\circ }\in \mathfrak{g}$, identifying the extended
Hilbert space $\mathfrak{g}=\mathbb{C}\oplus \mathfrak{f}$ \ with the space $%
L_{\mu \oplus 1}^{2}$ of square-integrable functions of all $y$ by the
extension $\mu \oplus 1$ of the measure $\mu $ at $y^{\circ }$ as \textrm{d}$%
\mu _{y^{\circ }}=1$.

Indeed, we can always assume that $V\left( y\right) =\mathrm{e}^{-i\mathrm{E}%
/\hbar }F\left( y\right) $ where the family $F$ is viewed as an isometry $%
\mathrm{F}:\mathfrak{h}\rightarrow \mathfrak{h}\otimes \mathfrak{f}$
corresponding $\mathrm{F}^{\dagger }\mathrm{F}=\mathrm{I}$ (not necessarily
of the form $F\left( y\right) =\chi ^{\circ }\left( y-\mathrm{X}\right) $ as
in (\ref{4.2})). Denoting $\mathrm{e}^{-i\mathrm{E}/\hbar }\mathrm{F}$ as
the column of $\mathrm{W}_{0}^{y}$, $y\neq 0$, and $\mathrm{e}^{-i\mathrm{E}%
/\hbar }\mathrm{F}^{\dagger }$ as the raw of $\mathrm{W}_{y}^{0}$, $y\neq 0$%
, we can compose the unitary block-matrix 
\begin{equation}
\left[ \mathrm{W}_{y^{\prime }}^{y}\right] :=\mathrm{e}^{-i\mathrm{E}/\hbar }%
\left[ 
\begin{tabular}{rr}
$\mathrm{O}$ & $\mathrm{F}^{\dagger }$ \\ 
$\mathrm{F}$ & $\mathrm{I}\otimes \hat{1}-\mathrm{FF}^{\dagger }$%
\end{tabular}
\right] ,\quad \mathrm{I}\otimes \hat{1}=\left[ \mathrm{I}\delta _{y^{\prime
}}^{y}\right] _{y^{\prime }\neq 0}^{y\neq 0}  \label{4.4}
\end{equation}
describing an operator $\mathrm{W}=\left[ \mathrm{W}_{y^{\prime }}^{y}\right]
$\ on the product $\mathfrak{h}\otimes \mathfrak{g}$, where $\mathfrak{g}=%
\mathbb{C\oplus }\mathfrak{f}$, represented as$\mathfrak{h}\oplus \left( 
\mathfrak{h}\otimes \mathfrak{f}\right) $, $\mathfrak{f}=L_{\mu }^{2}$. It
has the adjoint $\mathrm{W}^{\dagger }=\mathrm{e}^{i\mathrm{E}/\hbar }%
\mathrm{We}^{-i\mathrm{E}/\hbar }$, and obviously 
\begin{equation*}
\left( \mathrm{I}\otimes \langle y|\right) \mathrm{W}\left( \mathrm{I}%
\otimes |0\rangle \right) =V\left( y\right) ,\quad \forall y\neq 0.
\end{equation*}
The unitarity $\mathrm{W}^{-1}=\mathrm{W}^{\dagger }$ of the constructed
operator $\mathrm{W}$ is the consequence of the isometricity $\mathrm{F}%
^{\dagger }\mathrm{F}=\mathrm{I}$ and thus the projectivity $\left( \mathrm{%
FF}^{\dagger }\right) ^{2}=\mathrm{FF}^{\dagger }$ of $\mathrm{FF}^{\dagger
} $ and of $\mathrm{I}\otimes \hat{1}-\mathrm{FF}^{\dagger }$: 
\begin{equation*}
\mathrm{W}^{\dagger }\mathrm{W}=\left[ 
\begin{array}{ll}
\mathrm{F}^{\dagger }\mathrm{F} & \mathrm{F}^{\dagger }\left( \mathrm{I}%
\otimes \hat{1}-\mathrm{FF}^{\dagger }\right) \\ 
\left( \mathrm{I}\otimes \hat{1}-\mathrm{FF}^{\dagger }\right) \mathrm{F} & 
\mathrm{FF}^{\dagger }+\mathrm{I}\otimes \hat{1}-\mathrm{FF}^{\dagger }%
\end{array}
\right] =\left[ 
\begin{array}{ll}
\mathrm{I} & \mathrm{O} \\ 
\mathrm{O} & \mathrm{I}\otimes \hat{1}%
\end{array}
\right] .
\end{equation*}

In general the observation may be incomplete: the data $y$ may be the only
observable part of a pair $\left( z,y\right) $ defining the stochastic wave
propagator $V\left( z,y\right) .$ Consider for simplicity a discrete $z$
such that 
\begin{equation*}
V^{\dagger }V:=\sum_{z}\int V\left( z,y\right) ^{\dagger }V\left( z,y\right) 
\mathrm{d}\mu _{y}=\mathrm{I}.
\end{equation*}
Then the linear unital map on the algebra $\mathcal{B}\left( \mathfrak{h}%
\right) \otimes \mathcal{C}$ of the completely positive form 
\begin{equation*}
\pi \left( \hat{g}\mathrm{B}\right) =\sum_{z}\int g\left( y\right) V\left(
z,y\right) ^{\dagger }\mathrm{B}V\left( z,y\right) \mathrm{d}\mu _{y}\equiv 
\mathsf{M}\left[ g\pi \left( \mathrm{B}\right) \right]
\end{equation*}
describes the ''Heisenberg picture'' for generalized von Neumann reduction
with an incomplete measurement results $y$. Here $\mathrm{B}\in \mathcal{B}%
\left( \mathfrak{h}\right) $, $\hat{g}$ is the multiplication operator by a
measurable function of $y$ defining any system-pointer observable by linear
combinations of $B\left( y\right) =g\left( y\right) \mathrm{B}$, and 
\begin{equation*}
\pi \left( y,\mathrm{B}\right) =\sum_{z}V\left( z,y\right) ^{\dagger }%
\mathrm{B}V\left( z,y\right) ,\quad \mathsf{M}\left[ B\left( y\right) \right]
=\int B\left( y\right) \mathrm{d}\mu _{y}.
\end{equation*}
The function $y\mapsto \pi \left( y\right) $ with values in the completely
positive maps $\mathrm{B}\mapsto \pi \left( y,\mathrm{B}\right) $, or
operations, is the basic tool in the operational approach to quantum
measurements. Its adjoint 
\begin{equation*}
\pi ^{\ast }\left( \sigma \right) =\sum_{z}V\left( y,z\right) \sigma V\left(
y,z\right) ^{\dagger }\mathrm{d}\mu _{y}=\pi ^{\ast }\left( y,\sigma \right) 
\mathrm{d}\mu _{y},
\end{equation*}
is given by the density matrix transformation and it is called the \emph{%
instrument} in the phenomenological measurement theories. The operational
approach was introduced by Ludwig \cite{Lud68}, and the mathematical
implementation of the notion of instrument was originated by Davies and
Lewis \cite{DaLe70}.

An abstract instrument now is defined as the adjoint to a unital completely
positive map $\pi $ for which $\pi _{y}^{\ast }\left( \sigma \right) $ is a
trace-class operator for each $y$, normalized to a density operator $\rho
=\int \mathrm{d}\pi _{y}^{\ast }\left( \sigma \right) $. The quantum mixed
state described by the operator $\rho $ is called the \emph{prior} state,
i.e. the state which has been prepared for the measurement. A unitary
dilation of the generalized reduction (or ``instrumental'') map $\pi $ was
constructed by Ozawa \cite{Oz84}, but as we shall now see, this, as well as
the canonical dilation (\ref{4.4}), is only a preliminary step towards the
its quantum stochastic realization allowing the dynamical derivation of the
reduction postulate as a result of the statistical inference as it was
suggested in \cite{Be94}.

\section{The future-past boundary value problem}

The additional system of the constructed unitary dilation for the
measurement propagator $V\left( y\right) $ represents only the pointer
coordinate of the measurement apparatus $y$ with the initial value $%
y=y^{\circ }$ ($=0$ corresponding to $\chi ^{\circ }=|0\rangle $). It should
be regarded as a classical system (like the Schr\"{o}dinger's cat) at the
instants of measurement $t>0$ in order to avoid the applying of the
projection postulate for inferences in the auxiliary system. Indeed, the
actual events of the measurement can be only those propositions $E$ in the
extended system which may serve as the conditions for any other proposition $%
F$ as a potential in future event, otherwise there can't be any causality
even in the weak, statistical sense. This means that future states should be
statistically predictable in any prior state of the system in the result of
testing the measurable event $E$ by the usual conditional probability
(Bayes) formula 
\begin{equation}
\Pr \left\{ F=1|E=1\right\} =\Pr \left\{ E\wedge F=1\right\} /\Pr \left\{
E=1\right\} \text{ \quad }\forall F,  \label{4.5}
\end{equation}
and this predictability, or statistical causality means that the prior
quantum probability $\Pr \left\{ F\right\} \equiv \Pr \left\{ F=1\right\} $
must coincide with the statistical expectation of $F$ as\ the weighted sum 
\begin{equation*}
\Pr \left\{ F|E\right\} \Pr \left\{ E\right\} +\Pr \left\{ F|E^{\perp
}\right\} \Pr \left\{ E^{\perp }\right\} =\Pr \left\{ F\right\}
\end{equation*}
of this $\Pr \left\{ F|E\right\} \equiv \Pr \left\{ F=1|E=1\right\} $ and
the complementary conditional probability $\Pr \left\{ F|E^{\perp }\right\}
=\Pr \left\{ F=1|E=0\right\} $. As one can easily see, this is possible if
and only if (\ref{2.7}) holds, i.e. any other future event-orthoprojector $F$
of the extended system must be compatible with the actual
event-orthoprojector $E$.

The actual events in the measurement model obtained by the unitary dilation
are only the orthoprojectors $E=\mathrm{I}\otimes \hat{1}_{\Delta }$ on $%
\mathfrak{h}\otimes \mathfrak{g}$ corresponding to the propositions ''$y\in
\Delta $'' where $\hat{1}_{\Delta }$ is the multiplication by the indicator $%
1_{\Delta }$ for a measurable on the pointer scale subset $\Delta $. Other
orthoprojectors which are not compatible with these orthoprojectors, are
simply not admissible as the questions by the choice of time arrow. This
choice restores the quantum causality as statistical predictability, i.e.
the statistical inference made upon the sample data. And the actual
observables in question are only the measurable functions $g\left( y\right) $
of $y\neq y^{\circ }$ represented on $\mathfrak{f}=L^{2}\left( \mu \right) $
by the commuting operators $\hat{g}$ of multiplication by these functions, $%
\langle y|\hat{g}\chi =g\left( y\right) \chi \left( y\right) $. As follows
from $\mathrm{W}_{0}^{0}=\mathrm{O}$, the initial value $y^{\circ }=0$ is
never observed at the time $t=1$: 
\begin{equation*}
\left\| \psi _{1}\left( 0\right) \right\| ^{2}=\left\| \left( \mathrm{I}%
\otimes \langle 0|\right) \mathrm{W}\left( \psi \otimes |0\rangle \right)
\right\| ^{2}=\left\| \mathrm{W}_{0}^{0}\psi \right\| ^{2}=0,\quad \forall
\psi \in \mathfrak{h}
\end{equation*}
(that is a measurable value $y\neq y^{\circ }$ is certainly observed at $t=1$%
). These are the only appropriate candidates for Bell's ''beables'', \cite%
{Bell87}, p.174. Indeed, such commuting observables, extended to the quantum
counterpart as $\mathrm{G}_{0}=\mathrm{I}\otimes \hat{g}$ on $\mathfrak{h}%
\otimes \mathfrak{f}$, are compatible with any admissible question or
observable $\mathrm{B}$ on $\mathfrak{h}$ represented with respect to the
output states $\psi _{1}=\mathrm{W}\psi _{0}$ at the time of measurement $%
t=1 $ by an operator $\mathrm{B}_{1}=\mathrm{B}\otimes \hat{1}$ on $%
\mathfrak{h}\otimes \mathfrak{f}$. The probabilities (or, it is better to
say, the propensities) of all such questions are the same in all states
whether an observable $\mathrm{G}_{0}$ was measured but the result not read,
or it was not measured at all. In this sense the measurement of $\mathrm{G}%
_{0}$ is called \textit{nondemolition} with respect to the system
observables $\mathrm{B}_{1}$, they do not demolish the propensities, or
prior expectations of $\mathrm{B}$. However as we shall show now they are
not necessary compatible with the same operators $\mathrm{B}$ of the quantum
system at the initial stage and\ currently represented as $\mathrm{WB}_{0}%
\mathrm{W}^{\dagger }$ on $\psi _{1} $, where $\mathrm{B}_{0}=\mathrm{B}%
\otimes \hat{1}$ is the$\,$\ Schr\"{o}dinger representation of $\mathrm{B}$
at the time $t=0$ on the corresponding input states $\psi _{0}=\mathrm{W}%
^{\dagger }\psi _{1}$ in $\mathfrak{h}\otimes \mathfrak{g}$ $.$

Indeed, we can see this on the example of the Schr\"{o}dinger cat, where $%
\mathrm{W}$ is the flip $\mathrm{S}$ in $\mathfrak{g}=\mathbb{C}^{2}$ (shift 
$\func{mod}2$). In this case the operators the operators $\hat{g}_{1}$ in
the Heisenberg picture $\mathrm{G}=\mathrm{S}^{\dagger }\mathrm{G}_{0}%
\mathrm{S}$ are represented on $\mathfrak{h}\otimes \mathfrak{g}$ as the
diagonal operators $\mathrm{G}=\left[ g\left( \tau +\upsilon \right) \delta
_{\tau ^{\prime }}^{\tau }\delta _{\upsilon ^{\prime }}^{\upsilon }\right] $
of multiplication by $g\left( \tau +\upsilon \right) $, where the sum $\tau
+\upsilon =\left| \tau -\upsilon \right| $ is modulo 2. Obviously they do
not commute with $\mathrm{B}_{0}$ unless $\mathrm{B}$ is also a diagonal
operator $\hat{f}$ of multiplication by a function $f\left( \tau \right) $,
in which case 
\begin{equation*}
\left[ \mathrm{B}_{0},\mathrm{G}\right] \psi _{0}\left( \tau ,\upsilon
\right) =\left[ f\left( \tau \right) ,g\left( \tau +\upsilon \right) \right]
\psi _{0}\left( \tau ,\upsilon \right) =0,\quad \forall \psi _{0}\in 
\mathfrak{h}\otimes \mathfrak{g}\text{.}
\end{equation*}
The restriction of the possibilities in a quantum system to only the
diagonal operators $\mathrm{B}=\hat{f}$ of the atom which would eliminate
the time arrow in the nondemolition condition, amounts to the redundancy of
the quantum consideration: all such (possible and actual) observables can be
simultaneously represented as classical observables by the measurable
functions of $\left( \tau ,\upsilon \right) $.

Thus the constructed semiclassical algebra $\mathcal{B}_{-}=\mathcal{B}%
\left( \mathfrak{h}\right) \otimes \mathcal{C}$ of the Schr\"{o}dinger's
atom and the pointer (dead or alive cat) is not dynamically invariant in the
sense that transformed algebra $\mathrm{W}^{\dagger }\mathcal{B}_{-}\mathrm{W%
}$ does not coincide and is not a part of $\mathcal{B}_{-}$ but of $\mathcal{%
B}_{+}=\mathcal{B}\left( \mathfrak{h}\right) \otimes \mathcal{B}\left( 
\mathfrak{g}\right) $. This is also true in the general case, unless all the
system-pointer observables in the Heisenberg picture are still decomposable, 
\begin{equation*}
\mathrm{W}^{\dagger }\left( \mathrm{B}\otimes \hat{g}\right) \mathrm{W}=\int
|y\rangle g\left( y\right) B\left( y\right) \langle y|\mathrm{d}\mu _{y},
\end{equation*}
which would imply $\mathrm{W}^{\dagger }\mathcal{B}\mathrm{W}\subseteq 
\mathcal{B}$. (Such dynamical invariance of the decomposable algebra , given
by the operator-valued functions $B\left( y\right) $, can be achieved by
this unitary dilations only in trivial cases.) This is why the von Neumann
type dilation (\ref{3.1}), and even more general dilations (\ref{4.4}), or 
\cite{Oz84, Be94} cannot yet be considered as the dynamical solution of the
instantaneous quantum measurement problem which we formulate in the
following way.

\emph{Given a reduction postulate defined by an isometry }$V$ \emph{on} $%
\mathfrak{h}$ \emph{into }$\mathfrak{h}\otimes \mathfrak{g}$\emph{, find a
triple }$\left( \mathcal{G},\mathfrak{A},\Phi ^{\circ }\right) $ \emph{%
consisting of Hilbert space} $\mathcal{G=G}_{-}\otimes \mathcal{G}_{+}$ 
\emph{embedding the Hilbert spaces }$\mathfrak{f=}L_{\mu }^{2}$\emph{\ by an
isometry into }$\mathcal{G}_{+}$\emph{, an algebra }$\mathfrak{A=A}%
_{-}\otimes \mathfrak{A}_{+}$ \emph{on $\mathcal{G}$ with an Abelian
subalgebra }$\mathfrak{A}_{-}=\mathfrak{C}$\emph{\ generated by an
observable (beable) }$Y$\emph{\ on $\mathcal{G}_{-}$, and a state-vector }$%
\Phi ^{\circ }=\Phi _{-}^{\circ }\otimes \Phi _{+}^{\circ }\in \mathcal{G}$%
\emph{\ such that there exist a unitary operator }$U$\emph{\ on }$\mathcal{H=%
}\mathfrak{h}\otimes \mathcal{G}$\emph{\ which induces an endomorphism on
the product algebra $\mathfrak{B=}\mathcal{B}\left( \mathfrak{h}\right)
\otimes \mathfrak{A}$ in the sense }$U^{\dagger }AU\in \mathfrak{B}$ \emph{%
for all }$B\in \mathfrak{B}$\emph{, with } 
\begin{equation*}
\pi \left( \hat{g}\otimes \mathrm{B}\right) :=\left( \mathrm{I}\otimes \Phi
^{\circ }\right) ^{\dagger }U^{\dagger }\left( \mathrm{B}\otimes g\left(
Y\right) \right) U\left( \mathrm{I}\otimes \Phi ^{\circ }\right) =\mathsf{M}%
\left[ gV^{\dagger }\mathrm{B}V\right]
\end{equation*}
\emph{\ for all }$\mathrm{B}\in \mathcal{B}\left( \mathfrak{h}\right) $ 
\emph{and measurable functions }$g$\emph{\ of }$Y$\emph{, where }$\mathsf{M}%
\left[ B\right] =\int B\left( y\right) \mu _{y}$.

It is always possible to achieve this dynamical invariance by extending the
classical measurement apparatus' to an infinite auxiliary semi-classical
system. Here we sketch this construction for the general unitary dilation (%
\ref{4.4}).

The construction consists of five steps. The first, preliminary step of a
unitary dilation for the isometry $\mathrm{V}$ has been already described in
the previous Section.

Second, we construct the triple $\left( \mathcal{G},\mathfrak{A},\Phi
^{\circ }\right) $. Denote by $\mathfrak{g}_{s}$, $s=\pm 0,\pm 1,\ldots $
(the indices $\pm 0$ are distinct and ordered as $-0<+0$) the copies of the
Hilbert space $\mathfrak{g}=\mathbb{C}\oplus \mathfrak{f}$ in the dilation (%
\ref{4.4}) represented as the functional space $L_{\mu }^{2}$ on the values
of $y$ including $y^{\circ }=0$, and $\mathbb{G}_{n}=\mathfrak{g}%
_{-n}\otimes \mathfrak{g}_{+n}$, $n\geq 0$. We define the Hilbert space of
the past $\mathcal{G}_{-}$ and the future $\mathcal{G}_{+}$ as the
state-vector spaces of semifinite discrete strings generated by the infinite
tensor products $\Phi _{-}=\chi _{-0}\otimes \chi _{-1}\otimes \ldots $ and $%
\Phi _{+}=\chi _{+0}\otimes \chi _{+1}\otimes \ldots $ with all but finite
number of $\chi _{s}\in \mathfrak{g}_{s}$ equal to the initial state $\chi
_{s}^{\circ }$, the copies of $\chi ^{\circ }=$ $|0\rangle \in \mathfrak{g}$%
. Denoting by $\mathcal{A}_{s}$ the copies of the algebra $\mathcal{B}\left( 
\mathfrak{g}\right) $ of bounded operators if $s\geq +0$, of the diagonal
subalgebra $\mathcal{D}\left( \mathfrak{g}\right) $ on $\mathfrak{g}$ if $%
s\leq -0$, \ and $\mathcal{A}_{n}=\mathcal{A}_{-n}\otimes \mathcal{A}_{+n}$
we construct the algebras of the past $\mathfrak{A}_{-}$ and the future $%
\mathfrak{A}_{+}$ and the whole algebra $\mathfrak{A}$. $\mathfrak{A}_{\pm }$
are generated on $\mathcal{G}_{\pm }$ respectively by the diagonal operators 
$\hat{f}_{-0}\otimes \hat{f}_{-1}\otimes \ldots $ and by $\mathrm{X}%
_{+0}\otimes \mathrm{X}_{+1}\otimes \ldots $ with all but finite number of $%
\hat{f}_{s}\in \mathcal{A}_{s}$, $s<0$ and $\mathrm{X}_{s}\in \mathcal{A}%
_{s} $, $s>0$ equal the identity operator $\hat{1}$ in $\mathfrak{g}$. Here $%
\hat{f}$ stands for the multiplication operator by a function $f$ of $y\in 
\mathbb{R}$, in particular, $\hat{y}$ is the multiplication by $y$, with the
eigen--vector $\chi ^{\circ }=|0\rangle $ corresponding to the eigen-value $%
y^{\circ }=0$. The Hilbert space $\mathcal{G}_{-}\otimes \mathcal{G}_{+}$
identified with $\mathcal{G}=\otimes \mathbb{G}_{n}$, the decomposable
algebra $\mathfrak{A}_{-}\otimes \mathfrak{A}_{+}$ identified with $%
\mathfrak{A}=\otimes \mathcal{A}_{n}$ and the product vector $\Phi
_{-}\otimes \Phi _{+}$ identified with $\Phi =\otimes \phi _{n}\in \mathcal{G%
}$, where $\phi _{n}=\chi _{-n}\otimes \chi _{+n}\equiv \chi _{-n}\chi _{+n}$
with all $\chi _{s}=\chi ^{\circ }$ stand as candidates for the triple $%
\left( \mathcal{G},\mathfrak{A},\Phi \right) $. Note that the eigen-vector $%
\Phi ^{\circ }=\otimes \phi _{n}^{\circ }$ with all $\phi _{n}^{\circ }=\chi
^{\circ }\otimes \chi ^{\circ }$ corresponds to the initial eigen-state $%
y^{\circ }=0$ of all observables $Y_{\pm n}=\hat{1}_{0}\otimes \ldots
\otimes \hat{1}_{n-1}\otimes \hat{y}_{\pm }\otimes \hat{1}_{n+1}\otimes $ in 
$\mathcal{G}$, where $\hat{1}=\hat{1}_{-}\otimes \hat{1}_{+}$, $\hat{y}_{-}=%
\hat{y}\otimes \hat{1}_{+}$, $\hat{y}_{+}=\hat{1}_{-}\otimes \hat{y}$ and $%
\hat{1}_{\pm n}$ are the identity operators in $\mathfrak{g}_{\pm n}$.

Third, we define the unitary evolution on the product space $\mathfrak{h}%
\otimes \mathcal{G}$ of the total system by 
\begin{equation}
U:\psi \otimes \chi _{-}\chi _{+}\otimes \chi _{-1}\chi _{+1}\cdots \mapsto 
\mathrm{W}\left( \psi \otimes \chi _{+}\right) \chi _{+1}\otimes \chi
_{-}\chi _{+2}\cdots ,  \label{4.6}
\end{equation}
incorporating the right shift in $\mathcal{G}_{-}$, the left shift in $%
\mathcal{G}_{+}$ and the conservative boundary condition $\mathrm{W}:%
\mathfrak{h}\otimes \mathfrak{g}_{+}\rightarrow \mathfrak{h}\otimes 
\mathfrak{g}_{-}$ given by the unitary dilation (\ref{4.4}). We have
obviously 
\begin{equation*}
\left( \mathrm{I}\otimes \langle y_{-},y_{+},y_{-1},y_{+1}\ldots |\right)
U\left( \mathrm{I}\otimes |y_{-}^{0},0,y_{-1}^{0},0\ldots \rangle \right)
=\cdots \delta _{0}^{y_{+1}}\delta _{0}^{y_{+}}V\left( y_{-}\right) \delta
_{y_{-}^{0}}^{y_{-1}}\delta _{y_{-1}^{0}}^{y_{-2}}\cdots
\end{equation*}
so that the extended unitary operator $U$ still reproduces the reduction $%
V\left( y\right) $ in the result $y\neq y^{\circ }$ of the measurement $%
Y=Y_{-0}$ in a sequence $\left( Y_{-0},Y_{+0},Y_{-1}Y_{+1},\ldots \right) $
with all other $y_{s}$ being zero $y^{\circ }=0$ with the probability one
for the initial ground state $\Phi ^{\circ }$ of the connected string.

Fourth, we prove the dynamical invariance $U^{\dagger }\left( \mathcal{B}%
\left( \mathfrak{h}\right) \otimes \mathfrak{A}\right) U\subseteq \mathcal{B}%
\left( \mathfrak{h}\right) \otimes \mathfrak{A}$ of the decomposable algebra
of the total system, incorporating the measured quantum system $\mathcal{B}%
\left( \mathfrak{h}\right) $ as the boundary between the quantum future (the
right string considered as quantum, $\mathfrak{A}_{+}=\mathcal{B}\left( 
\mathcal{G}_{+}\right) $) with the classical past (the left string
considered as classical, $\mathfrak{A}_{-}=\mathcal{D}\left( \mathcal{G}%
_{-}\right) $ ). This follows straightforward from the definition of $U$%
\begin{equation*}
U^{\dagger }\left( \mathrm{B}\otimes \hat{g}_{-}\mathrm{X}_{+}\otimes \hat{g}%
_{-1}\mathrm{X}_{+1}\cdots \right) U=\hat{g}_{-1}\mathrm{W}^{\dagger }\left( 
\mathrm{B}\otimes \hat{g}_{-}\right) \mathrm{W}\otimes \hat{g}_{-2}\mathrm{X}%
_{+}\cdots
\end{equation*}
due to $\mathrm{W}^{\dagger }\left( \mathrm{B}\otimes \hat{g}\right) \mathrm{%
W}\in \mathcal{B}\left( \mathfrak{h}\right) \otimes \mathcal{B}\left( 
\mathfrak{g}\right) $ for all $\hat{g}\in \mathcal{D}\left( \mathfrak{g}%
\right) $. However this algebra representing the total algebra $\mathcal{B}%
\left( \mathfrak{h}\right) \otimes \mathfrak{A}$ on $\mathfrak{h}\otimes 
\mathcal{G}$ is not invariant under the inverse transformation, and there in
no way to achieve the inverse invariance keeping $\mathfrak{A}$ decomposable
as the requirement for statistical causality of quantum measurement if $%
\mathrm{W}\left( \mathrm{B}\otimes \mathrm{X}\right) \mathrm{W}^{\dagger
}\notin \mathcal{B}\left( \mathfrak{h}\right) \otimes \mathcal{D}\left( 
\mathfrak{g}\right) $ for some $\mathrm{B}\in \mathcal{B}\left( \mathfrak{h}%
\right) $ and $\mathrm{X}\in \mathcal{B}\left( \mathfrak{g}\right) $: 
\begin{equation*}
U\left( \mathrm{B}\otimes \hat{g}_{-}\mathrm{X}_{+}\otimes \hat{g}_{-1}%
\mathrm{X}_{+1}\cdots \right) U^{\dagger }=\mathrm{W}\left( \mathrm{B}%
\otimes \mathrm{X}_{+}\right) \mathrm{W}^{\dagger }\mathrm{X}_{+1}\otimes 
\hat{g}_{-}\mathrm{X}_{+2}\cdots .
\end{equation*}

And the fifth step is to explain on this dynamical model the decoherence
phenomenon, irreversibility and causality by giving a constructive scheme in
terms of equation for quantum predictions as statistical inferences by
virtue of gaining the measurement information.

Because of the crucial importance of these realizations for developing
understanding of the mathematical structure and interpretation of modern
quantum theory, we need to analyze the mathematical consequences which can
be drawn from such schemes.

\section{Decoherence and quantum prediction}

The analysis above shows that the dynamical realization of a quantum
instantaneous measurement is possible in an infinitely extended system, but
the discrete unitary group of unitary transformations $U^{t}$, $t\in \mathbb{%
N}$ with $U^{1}=U$ induces not a group of Heisenberg authomorphisms but an
injective irreversible semigroup of endomorphisms on the decomposable
algebra $\mathfrak{B}=\mathcal{B}\left( \mathfrak{h}\right) \otimes 
\mathfrak{A}$ of this system. However it is locally invertible on the center
of the algebra $\mathfrak{A}$ in the sense that it reverses the shift
dynamics on $\mathfrak{A}^{0]}$: 
\begin{equation}
T_{-t}\left( \mathrm{I}\otimes Y_{s}\right) T_{t}:=\mathrm{I}\otimes
Y_{s-t}=U^{t}\left( \mathrm{I}\otimes Y_{s}\right) U^{-t},\quad \forall
s\leq -0,t\in \mathbb{N}.  \label{4.7}
\end{equation}
Here $Y_{-n}=\hat{1}^{\otimes n}\otimes \hat{y}_{-}\otimes I_{n}$, where $%
I_{n}=\otimes _{k>n}\hat{1}_{k}$, and $T_{-t}=\left( T\right) ^{t}$ is the
power of the isometric shift $T:\Phi _{-}\mapsto \chi ^{\circ }\otimes \Phi
_{-}$ on $\mathcal{G}_{-}$ extended to the free unitary dynamics of the
whole system as 
\begin{equation*}
T:\psi \otimes \chi _{-}\chi _{+}\otimes \chi _{-1}\chi _{+1}\cdots \mapsto
\psi \otimes \chi _{+}\chi _{+1}\otimes \chi _{-}\chi _{+2}\cdots .
\end{equation*}

The extended algebra $\mathfrak{B}$ is the minimal algebra containing all
consistent events of the history and all admissible questions about the
future of the open system under observation initially described by $\mathcal{%
B}\left( \mathfrak{h}\right) $. Indeed, it contains all Heisenberg operators 
\begin{equation*}
B\left( t\right) =U^{-t}\left( \mathrm{B}\otimes I\right) U^{t},\quad
Y_{-}\left( t\right) =U^{-t}\left( \mathrm{I}\otimes Y_{-0}\right)
U^{t},\quad \forall t>0
\end{equation*}
of $\mathrm{B}\in \mathcal{B}\left( \mathfrak{h}\right) $, and these
operators not only commute at each $t$, but also satisfy \emph{the
nondemolition causality condition} 
\begin{equation}
\left[ B\left( t\right) ,Y_{-}\left( r\right) \right] =0,\quad \left[
Y_{-}\left( t\right) ,Y_{-}\left( r\right) \right] =0,\quad \forall t\geq
r\geq 0.  \label{4.8}
\end{equation}
This follows from the commutativity of the Heisenberg string operators 
\begin{equation*}
Y_{r-t}\left( t\right) =U^{-t}\left( \mathrm{I}\otimes Y_{r-t}\right)
U^{t}=Y_{-}\left( r\right)
\end{equation*}
at the different points $s=r-t<0$ coinciding with $Y_{s}\left( r-s\right) $
for any $s<0$ because of (\ref{4.7}), and also from the commutativity with $%
\mathrm{B}\left( t\right) $ due to the simultaneous commutativity of all $%
Y_{s}\left( 0\right) =\mathrm{I}\otimes \mathrm{Y}_{s}$ and $B\left(
0\right) =\mathrm{B}\otimes I$. Thus all output Heisenberg operators $%
Y_{-}\left( r\right) $, $0<r\leq t$ at the boundary of the string can be
measured simultaneously as $Y_{-n}\left( t\right) =Y_{-}\left( t-n\right) $
at the different points $n<t$, or sequentially at the point $s=-0$ as the
commutative nondemolition family $Y_{0}^{t]}=\left( Y^{1},\ldots
,Y^{t}\right) $, where $Y^{r}=Y_{-}\left( r\right) $. This defines the
reduced evolution operators 
\begin{equation*}
V\left( t,y_{0}^{t]}\right) =V\left( y^{t}\right) V\left( y^{t-1}\right)
\cdots V\left( y^{1}\right) ,\quad t>0
\end{equation*}
of a \emph{sequential measurement} in the system Hilbert space $\mathfrak{h}$
with measurement data $y_{0}^{t]}=\left\{ (0,t]\ni r\mapsto y^{r}\right\} $.
One can prove this using the \emph{filtering recurrency equation} 
\begin{equation}
\psi \left( t,y_{0}^{t]}\right) =V\left( y_{t}\right) \psi \left(
t-1,y_{0}^{t-1]}\right) ,\quad \psi \left( 0\right) =\psi  \label{4.9}
\end{equation}
for $\psi \left( t,y_{0}^{t]}\right) =V\left( t,y_{0}^{t]}\right) \psi $ and
for $\Psi \left( t\right) =U^{t}\left( \psi \otimes \Phi _{-}\otimes \Phi
_{+}^{\circ }\right) $, where $\psi \in \mathfrak{h}$, and $V\left(
y_{t}\right) $ is defined by 
\begin{equation*}
\left( \mathrm{I}\otimes \langle y_{-\infty }^{t]}|\otimes \langle
y_{t}^{\infty }|\right) U\Psi \left( t-1\right) =V\left( y_{t}\right) \psi
\left( t-1,y_{0}^{t-1]}\right) \langle \delta _{0}^{y_{t}^{\infty
}}y_{-\infty }^{0]}|\Phi _{-}.
\end{equation*}

Moreover, any future expectations in the system, say the probabilities of
the questions $F\left( t\right) =U^{-t}\left( F\otimes I\right) U^{t},t\geq
s $ \ given by orthoprojectors $F$ on $\mathfrak{h}$, can be statistically
predicted upon the results of the past measurements of $Y_{-}\left( r\right) 
$, $0<r\leq t$ and initial state $\psi $ by\ the simple conditioning 
\begin{equation*}
\Pr \left\{ F\left( t\right) |E\left( \mathrm{d}y^{1}\times \cdots \times 
\mathrm{d}y^{t}\right) \right\} =\frac{\Pr \left\{ F\left( t\right) \wedge
E\left( \mathrm{d}y^{1}\times \cdots \times \mathrm{d}y^{t}\right) \right\} 
}{\Pr \left\{ E\left( \mathrm{d}y^{1}\times \cdots \times \mathrm{d}%
y^{t}\right) \right\} }.
\end{equation*}
Here $E$ is the joint spectral measure for $Y^{1},\ldots ,Y^{t}$, and the
probabilities in the numerator (and denominator) are defined as 
\begin{equation*}
\left\| F\left( t\right) E\left( \mathrm{d}y^{1}\times \cdots \times \mathrm{%
d}y^{t}\right) \left( \psi \otimes \Phi ^{\circ }\right) \right\|
^{2}=\left\| F\psi \left( t,y_{0}^{t]}\right) \right\| ^{2}\mathrm{d}\mu
_{y^{1}}\cdots \mathrm{d}\mu _{y^{t}}
\end{equation*}
(and for $F=\mathrm{I}$) due to the commutativity of $F\left( t\right) $
with $E_{-}$. This implies the usual sequential instrumental formula 
\begin{equation*}
\left\langle \mathrm{B}\right\rangle \left( t,y_{0}^{r]}\right) =\frac{\psi
^{\dagger }\pi \left( t,y_{0}^{r]},\mathrm{B}\right) \psi }{\psi ^{\dagger
}\pi \left( t,y_{0}^{r]},\mathrm{I}\right) \psi }=\mathsf{M}\left[ \psi
_{y_{0}^{t]}}^{\dagger }\left( t\right) \mathrm{B}\psi _{y_{0}^{t]}}\left(
t\right) |y_{0}^{r]}\right]
\end{equation*}
for the future expectations of $\mathrm{B}\left( t\right) $ conditioned by $%
Y_{-}\left( 1\right) =y^{1},\ldots ,Y_{-}\left( s\right) =y^{s}$ for any $%
t>r $. Here $\psi _{y_{0}^{t]}}\left( t\right) =\psi \left(
t,y_{0}^{t]}\right) /\left\| \psi \left( t,y_{0}^{t]}\right) \right\| $, and 
\begin{equation*}
\pi \left( t,y_{0}^{r]},\mathrm{B}\right) =\idotsint V\left(
t,y_{0}^{r]},y_{r}^{t]}\right) ^{\dagger }\mathrm{B}V\left(
t,y_{0}^{r]},y_{r}^{t]}\right) \mathrm{d}\mu _{y^{_{r+1}}}\cdots \mathrm{d}%
\mu _{y^{t}}
\end{equation*}
is the sequential reduction map $V\left( t,y_{0}^{t]}\right) ^{\dagger }%
\mathrm{B}V\left( t,y_{0}^{t]}\right) $ defining the prior probability
distribution 
\begin{equation*}
\mathsf{P}\left( \mathrm{d}y_{0}^{t]}\right) =\psi ^{\dagger }\pi \left(
t,y_{0}^{t]},\mathrm{I}\right) \psi \mathrm{d}\mu _{y_{0}^{t]}}=\left\| \psi
\left( t,y_{0}^{t]}\right) \right\| ^{2}\mathrm{d}\mu _{y^{1}}\cdots \mathrm{%
d}\mu _{y^{t}}
\end{equation*}
integrated over $y_{r}^{t]}$ if these data are ignored for the quantum
prediction of the state at the time $t>r$.\ 

Note that the stochastic vector $\psi \left( t,y_{0}^{t]}\right) $,
normalized as 
\begin{equation*}
\idotsint \left\| \psi \left( t,y_{0}^{t]}\right) \right\| ^{2}\mathrm{d}\mu
_{y^{1}}\cdots \mathrm{d}\mu _{y^{t}}=1
\end{equation*}
depends linearly on the initial state vector $\psi \in \mathfrak{h}$.
However \emph{the posterior state vector} $\psi _{y_{0}^{t]}}\left( t\right) 
$ is nonlinear, satisfying the \emph{nonlinear stochastic recurrency equation%
} 
\begin{equation}
\psi _{y_{0}^{t]}}\left( t\right) =V_{y_{0}^{t-1]}}\left( t,y^{t}\right)
\psi _{y_{0}^{t-1]}}\left( t-1\right) ,\quad \psi \left( 0\right) =\psi ,
\label{4.10}
\end{equation}
where $V_{y_{0}^{t-1]}}\left( t,y^{t}\right) =\left\| V\left(
t-1,y_{0}^{t-1]}\right) \psi \right\| V\left( y^{t-1]}\right) /\left\|
V\left( t,y_{0}^{t]}\right) \psi \right\| $.

In particular one can always realize in this way any sequential observation
of the noncommuting operators $\mathrm{B}_{t}=\mathrm{e}^{i\mathrm{E}/\hbar }%
\mathrm{B}_{0}\mathrm{e}^{-i\mathrm{E}/\hbar }$ given by a selfadjoint
operator $\mathrm{B}_{0}$ with discrete spectrum and the energy operator $%
\mathrm{E}$ in $\mathfrak{h}$. It corresponds to the sequential collapse
given by $V\left( y\right) =\delta _{0}^{\mathrm{B}_{0}}\mathrm{e}^{-i%
\mathrm{E}/\hbar }$. Our construction suggests that any demolition
sequential measurement can be realized as the nondemolition by the
commutative family $Y_{-}\left( t\right) $, $t>0$ with a common eigenvector $%
\Phi ^{\circ }$ as the pointers initial state, satisfying the causality
condition (\ref{4.8}) with respect to all future Heisenberg operators $%
\mathrm{B}\left( t\right) .$ And the sequential collapse (\ref{4.10})
follows from the usual Bayes formula for conditioning of the compatible
observables due to the classical inference in the extended system. Thus, we
have solved the \emph{sequential quantum measurement problem} which can
rigorously be formulated as

\emph{Given a sequential reduction family }$V\left( t,y_{0}^{t]}\right)
,t\in \mathbb{N}$ \emph{of isometries resolving the filtering equation (\ref%
{4.9}) on} $\mathfrak{h}$\emph{\ into $\mathfrak{h}\otimes \mathfrak{f}%
^{\otimes t}$, find a triple }$\left( \mathcal{G},\mathfrak{A},\Phi \right) $
\emph{consisting of a Hilbert space} $\mathcal{G=G}_{-}\otimes \mathcal{G}%
_{+}$ \emph{embedding all tensor products }$\mathfrak{f}^{\otimes t}$\emph{\
of the Hilbert spaces }$\mathfrak{f=}L_{\mu }^{2}$\emph{\ by an isometry
into }$\mathcal{G}_{+}$\emph{, an algebra }$\mathfrak{A=A}_{-}\otimes 
\mathfrak{A}_{+}$ \emph{on $\mathcal{G}$ with an Abelian subalgebra }$%
\mathfrak{A}_{-}=\mathfrak{C}$\emph{\ generated by a compatible discrete
family }$Y_{-\infty }^{0]}=\left\{ Y_{s}\text{ }s\leq 0\right\} $\emph{\ } 
\emph{of the observables (beables) }$Y_{s}$\emph{\ on $\mathcal{G}_{-}$, and
a state-vector }$\Phi ^{\circ }=\Phi _{-}^{\circ }\otimes \Phi _{+}^{\circ
}\in \mathcal{G}$\emph{\ such that there exist a unitary group }$U^{t}$\emph{%
\ on }$\mathcal{H=}\mathfrak{h}\otimes \mathcal{G}$\emph{\ inducing a
semigroup of endomorphisms }$\mathfrak{B}\ni B\mapsto U^{-t}BU^{t}\in 
\mathfrak{B}$ \emph{on the product algebra $\mathfrak{B=}\mathcal{B}\left( 
\mathfrak{h}\right) \otimes \mathfrak{A}$ (\ref{4.7}on }$\mathfrak{A}$\emph{%
, with } 
\begin{equation*}
\pi ^{t}\left( \hat{g}_{-t}\otimes \mathrm{B}\right) =\left( \mathrm{I}%
\otimes \Phi ^{\circ }\right) ^{\dagger }U^{-t}\left( g_{-t}\left(
Y_{-t}^{0]}\right) \otimes \mathrm{B}\right) U^{t}\left( \mathrm{I}\otimes
\Phi ^{\circ }\right) =\mathsf{M}\left[ gV\left( t\right) ^{\dagger }\mathrm{%
B}V\left( t\right) \right]
\end{equation*}
\emph{\ for any }$\mathrm{B}\in \mathcal{B}\left( \mathfrak{h}\right) $ 
\emph{and any operator }$\hat{g}_{-t}=\hat{g}_{-t}\left( Y_{-t}^{0]}\right)
\in \mathfrak{C}$ $\ $\emph{represented as the shifted function }$\hat{g}%
_{-t}\left( y_{-t}^{0]}\right) =g\left( y_{0}^{t]}\right) $\emph{\ of }$%
Y_{-t}^{-0]}=\left( Y_{1-t},\ldots ,Y_{0}\right) $ \emph{on $\mathcal{G}$ by
any measurable function }$g$\emph{\ of }$y_{0}^{t]}=\left( y_{1},\ldots
,y_{t}\right) $\emph{\ with arbitrary }$t>0$\emph{, where} 
\begin{equation*}
\mathsf{M}\left[ gV\left( t\right) ^{\dagger }\mathrm{B}V\left( t\right) %
\right] =\idotsint g\left( y_{0}^{t]}\right) V\left( t,y_{0}^{t]}\right)
^{\dagger }\mathrm{B}V\left( t,y_{0}^{t]}\right) \mathrm{d}\mu
_{y^{1}}\cdots \mathrm{d}\mu _{y^{t}}.
\end{equation*}

Note that our construction of the solution to this problem admits also the
time reversed representation of the sequential measurement process described
by the isometry $\mathrm{V}$. The reversed system leaves in the same Hilbert
space, with the same initial state-vector $\Phi ^{\circ }$ in the auxiliary
space G, however the reversed auxiliary system is described by the reflected
algebra $\widetilde{\mathfrak{A}}=R\mathfrak{A}R$ where the reflection $R$
is described by the unitary flip-operator $R:\Phi _{-}\otimes \Phi
_{+}\mapsto \Phi _{+}\otimes \Phi _{-}$ on $\mathcal{G}=\mathcal{G}%
_{-}\otimes \mathcal{G}_{+}$. The past and future in the reflected algebra $%
\widetilde{\mathfrak{A}}=\mathfrak{A}_{+}\otimes \mathfrak{A}_{-}$ are
flipped such that its left subalgebra consists now of all operators on $%
\mathcal{G}_{-}$, $\widetilde{\mathfrak{A}}_{-}=\mathcal{B}\left( \mathcal{G}%
\right) \supset \mathfrak{A}_{-}$ and its right subalgebra is the diagonal
algebra $\widetilde{\mathfrak{A}}_{+}=\mathcal{D}\left( \mathcal{G}%
_{+}\right) \subset \mathfrak{A}_{+}$ on $\mathcal{G}_{+}$. The inverse
operators $U^{t},t<0$ induce the reversed dynamical semigroup of the
injective endomorphisms $B\mapsto U^{-t}BU^{t}$ which leaves invariant the
algebra $\widetilde{\mathfrak{B}}=\mathcal{B}\left( \mathfrak{h}\right)
\otimes \widetilde{\mathfrak{A}}$ but not $\mathfrak{B}$. The reversed
canonical measurement process is described by another family $%
Y_{[+0}^{\infty }=\left( Y_{+t}\right) $ of commuting operators $%
Y_{+t}=RY_{-t}R$ in $\widetilde{\mathfrak{A}}_{+}$, and the Heisenberg
operators 
\begin{equation*}
Y_{+}\left( t\right) =Y_{s}\left( t-s\right) =RY_{-}\left( -t\right) R,\quad
t<0,s>0,
\end{equation*}
are compatible and satisfy the reversed causality condition 
\begin{equation*}
\left[ B\left( t\right) ,Y_{+}\left( r\right) \right] =0,,\quad \left[
Y_{+}\left( t\right) ,Y_{+}\left( r\right) \right] =0,\quad \forall t\leq
r\leq 0.
\end{equation*}
It reproduces another, reversed sequence of the successive measurements 
\begin{equation*}
V^{\ast }\left( t,y_{[t}^{0}\right) =V^{\ast }\left( y_{t}\right) V^{\ast
}\left( y_{t+1}\right) \cdots V^{\ast }\left( y_{-1}\right) ,\quad t<0,
\end{equation*}
where $V^{\ast }\left( y\right) =\left( \mathrm{I}\otimes \langle y|\right) 
\mathrm{W}^{-1}\left( \mathrm{I}\otimes |0\rangle \right) $ depends on the
choice of the unitary dilation $\mathrm{W}$ of $\mathrm{V}$. In the case of
the canonical dilation (\ref{4.4}) uniquely defined up to the system
evolution between the measurements, we obtain $V^{\ast }\left( y\right)
=F\left( y\right) \mathrm{e}^{i\mathrm{E}/\hbar }$. If the system the
Hamiltonian is time-symmetric, i.e. $\overline{\mathrm{E}}=\mathrm{E}$ in
the sense $\mathrm{E}\bar{\psi}=\overline{\mathrm{E}\psi }$ with respect to
the complex (or another) conjugation in $\mathfrak{h}$, and if $\overline{F}%
\left( y\right) =\mathrm{e}^{-i\mathrm{E}/\hbar }F\left( \tilde{y}\right) 
\mathrm{e}^{i\mathrm{E}/\hbar }$, where $y\mapsto \tilde{y}$ is a covariant
flip, $\widetilde{\tilde{y}}=y$ (e.g. $\tilde{y}=y$, or reflection of the
measurement data under the time reflection $t\mapsto -t$), then $V^{\ast
}\left( y\right) =\overline{V}\left( \tilde{y}\right) $. This means that the
reversed measurement process can be described as time-reflected direct
measurement process under the $\ast $-conjugation $\psi ^{\ast }\left(
y\right) =\bar{\psi}\left( \tilde{y}\right) $ in the space $\mathfrak{h}%
\otimes \mathfrak{f}$. And it can be modelled as the time reflected direct
nondemolition process under the involution $J\left( \psi \otimes \Phi
\right) =\bar{\psi}\otimes R\Phi ^{\ast }$ induced by $\chi ^{\ast }\left(
y\right) =\bar{\chi}\left( \tilde{y}\right) $ in $\mathfrak{g}$ with the
flip-invariant eigen-value $y^{\circ }=0$ and $|0\rangle ^{\ast }=|0\rangle $
corresponding to the real ground state $\chi ^{\circ }\left( y\right)
=\delta _{y}^{0}$.

Thus, the choice of time arrow, which is absolutely necessary for restoring
statistical causality in quantum theory, is equivalent to a superselection
rule. This corresponds to a choice of the minimal algebra $\mathfrak{B}%
\subset \mathcal{B}\left( \mathcal{H}\right) $ generated by all admissible
questions on a suitable Hilbert space $\mathcal{H}$ of the nondemolition
representation for a process of the successive measurements. All consistent
events should be drown from the center of $\mathfrak{B}$: the events must be
compatible with the questions, otherwise the propensities for the future
cannot be inferred from the testing of the past. The decoherence is
dynamically induced by a unitary evolution from any pure state on the
algebra $\mathfrak{B}$ \ corresponding to the initial eigen-state for the
measurement apparatus pointer which is described by the center of $\mathfrak{%
B}$. Moreover, the reversion of the time arrow corresponds to another choice
of the admissible algebra. It can be implemented by a complex conjugation $J$
on $\mathcal{H}$ on the transposed algebra $\widetilde{\mathfrak{B}}=J%
\mathfrak{B}J$ . Note that the direct and reversed dynamics respectively on $%
\mathfrak{B}$ and on $\widetilde{\mathfrak{B}}$ are only endomorphic, and
that the invertible authomorphic dynamics induced on the total algebra $%
\mathcal{B}\left( \mathcal{H}\right) =\mathfrak{B}\vee \widetilde{\mathfrak{B%
}}$ does not reproduce the decoherence due to the redundancy of one of its
part for a given time arrow $t$.

The constructed dynamical realization of the instantaneous and sequential
measurements is the simple discrete-time analog of the solution to the
continuous boundary-value problem for quantum stochastic models of the
nondemolition measurements. This boundary value problem, which was obtained
recently by second quantization of the Dirac-type boundary value problem 
\cite{Be01a} for wave propagation on $\mathbb{R}_{+}$, gives an
implementation of an old idea of Schr\"{o}dinger \cite{Schr31} that the
quantum jumps and measurements should be derived from a boundary value
problem for ``waves from future'' interacting with the opposite ``messages
from the past''. This also gives a simple exactly solvable model in line
with more recent attempts of the transactional interpretation of quantum
mechanics \cite{Crm86}. The superselection causality principle which enables
such purely dynamical interpretation for quantum measurements allows only
the present and future to be quantum, defining the past as classical, stored
in the trajectories of the particles. As Lawrence Bragg, a Nobel prize
winner, once said, everything in the future is a wave, everything in the
past is a particle.

\end{document}